%% file: main.tex
\pdfoutput=1
\documentclass[11pt]{article}
\usepackage[utf8]{inputenc}
\usepackage[T1]{fontenc}
\usepackage{lmodern}
\usepackage[margin=1.1in]{geometry}
\usepackage{amsmath,amssymb}
\usepackage[hidelinks]{hyperref}
\makeatletter\g@addto@macro{\UrlBreaks}{\do\a\do\b\do\c\do\d\do\e\do\f\do\g\do\h\do\i\do\j\do\k\do\l\do\m\do\n\do\o\do\p\do\q\do\r\do\s\do\t\do\u\do\v\do\w\do\x\do\y\do\z\do\0\do\1\do\2\do\3\do\4\do\5\do\6\do\7\do\8\do\9}\makeatother
\usepackage[numbers,sort&compress]{natbib}
\usepackage{microtype}
\usepackage{booktabs}

\title{The Referee Was Right, and So Was Lean:\\ Proof, Peer Review, and the Machine, 1665--2026}
\author{Hiroki Fukui\thanks{Forensic psychiatrist; independent researcher. \texttt{fukui@somec.org}.
This paper has not been refereed. That is part of its argument.}}
\date{November 2026}

\begin{document}
\maketitle

\begin{abstract}
\input{sections/abstract}
\end{abstract}

\input{sections/00_prologue}
\input{sections/10_part1}
\input{sections/20_part2}
\input{sections/30_part3}
\input{sections/40_coda}

\bibliographystyle{plainnat}
\sloppy
\bibliography{refs}
\end{document}

%% file: sections/abstract.tex
Mathematics has repeatedly rebuilt the way it decides that a proof is correct: by correspondence and public challenge, by academies and journals, by refereeing, and by the slow consensus of a community. On 20 November 2026 a proof of the Riemann Hypothesis was released as seventeen million lines of Lean, written by Levent Alpöge with Claude and accompanied by a four-page note rather than a paper. Within a day it had been certified correct by a public process, without journal refereeing. This paper reviews the history of how proofs have been accepted, from the seventeenth century to the first machine-checked theorems, and argues that November 2026 marks a turning point. Once a statement and its assumptions have been fixed, a kernel can check whether a submitted proof establishes that statement; for problems that can be stated in this way, that check no longer needs peer review. The judgements it used to make do not disappear; they move to two other places: before the proof, in the choice of the formal statement, and after it, in the work of understanding why the theorem is true. I set beside the Riemann Hypothesis two cases: a proof that was refereed and published but not accepted, and my own proof of a famous problem about magic squares, which a referee declined on grounds of plausibility and in which formalisation in Lean later exposed a gap. The referee was right, and so was Lean, in different senses. The referee could not say where the error was, but named the one method by which I could find out; following it, I identified for myself what my manuscript had not proved. I write as an outsider to the profession, a forensic psychiatrist working with AI and Lean. In my own case the change removed any excuse that I went unread because of who I am; whether it has removed the effect of affiliation from mathematics as a whole, my case cannot show.

%% file: sections/00_prologue.tex
\section*{Prologue: a letter}

In 2026 I submitted a long paper to a number theory journal. It claimed to settle a well-known open problem: that there is no $3\times 3$ magic square whose nine entries are pairwise distinct positive squares. The paper came back without being sent out for a full review, with a short note from a reviewer. I have kept it ever since, because it turned out to be the most useful thing anyone has written about my mathematics.

The reviewer did not look for an error and explained, quite reasonably, why not: claimed proofs of this problem almost always hide a subtle mistake that is hard to find in a long paper, few number theorists would agree to search for it, and mine was very unlikely to be the exception. The advice was that, if I truly believed the argument, I should formalise it in Lean, which the reviewer thought feasible with the help of AI coding agents, and make the formal proof public before submitting it anywhere.

I took the advice literally. I had never written a line of Lean. I am a psychiatrist, not a mathematician; my day job is the assessment and treatment of offenders. With AI coding agents doing most of the typing, I spent the following months formalising the argument. Lean checked the reductions one by one. Formalising the last step forced me to state, as a separate claim, something my manuscript had passed over; once that claim stood on its own, I could see that the manuscript supplied no proof of it. It is the remaining condition on which my whole nonexistence argument depended. The reviewer's suspicion was right.

A few months later, on 20 November 2026, Levent Alpöge released a proof of the Riemann Hypothesis and of its generalisation to Dirichlet $L$-functions. He released no paper. He released a four-page note and seventeen million lines of Lean, written with Claude. Within a day, dozens of people had rebuilt the proof on their own machines, several independently written proof checkers had accepted it, and the Lean FRO and the Mathlib Initiative had issued a joint statement certifying that both the statement and the proof are correct \citep{leanfro2026statement}. No journal was involved. The proof was certified without journal refereeing.

The two events are the same event at different scales. In both, a human gatekeeper's judgement about whether a proof is correct was replaced by a machine's check. In my case the check showed exactly where the proof stopped; in Alpöge's it went all the way. This paper is about what that replacement means for the institution that mathematics has used, for a little over a century, to decide what is true: peer review.

My claim is simple, and I will try to state it more sharply than someone inside the academy would find comfortable. Once a statement and its assumptions have been fixed, a kernel can check whether a submitted proof establishes that statement. For problems that can be stated in this way, peer review is no longer what performs that check. It never fully did; it now does not need to. The trust has not vanished. It has moved: to the statement, to the library of definitions, and to a small kernel that anyone can re-run. What survives of refereeing moves to two other places: \emph{before} the proof, where someone must decide what the formal statement should say, and \emph{after} it, where someone must understand why it is true. The first is now the real review. The second is the real scholarship.

Part~I reviews how mathematics has decided that a proof is correct, from letters and challenges in the seventeenth century to journals, refereeing, community consensus, and the first machine-checked proofs. Part~II describes the turning point of November 2026 and sets beside it two cases: a proof that was refereed and published yet not accepted, and my own failed proof, whose gap formalisation exposed. Part~III takes apart the functions that refereeing has bundled together, asks which of them a proof checker performs, and asks what is left for people.

%% file: sections/10_part1.tex
\part*{Part I. How proofs have been accepted}
\addcontentsline{toc}{part}{Part I. How proofs have been accepted}

The belief that a mathematical result becomes true when it is published in a refereed journal is younger than most mathematicians assume. This part reviews, briefly and with no claim to completeness, the successive ways in which mathematicians have decided that a proof is correct. Each was a technology of trust suited to its time. None was simply ``reading the proof carefully''.

\section{Challenges, letters, and priority (1500--1700)}

In the sixteenth century a solution to an equation was something one kept. Scipione del Ferro solved the depressed cubic around 1515 and told almost no one \citep{mactutor_tartaglia}. In 1535 Antonio Fior, who had the method from del Ferro, challenged Niccolò Tartaglia to a public contest of thirty problems; Tartaglia found the general method in February 1535 and solved them all, while Fior solved none \citep{mactutor_tartaglia}. Correctness was demonstrated by performance before witnesses, not by inspection of an argument. When Cardano published the method in \emph{Ars Magna} in 1545, after obtaining it from Tartaglia under a promise of secrecy, the quarrel that followed was conducted in printed challenges and ended in a public disputation in Milan in 1548 \citep{feldmann_cardano}. What was being decided was less whether the rule was right than whose it was.

In the seventeenth century the centre of mathematical exchange was a correspondence network. Marin Mersenne, from the 1620s until his death in 1648, relayed letters, problems, and challenges among the mathematicians of Europe \citep{goldstein2013routine}. Challenges in his circle were a routine institution: a correspondent who could solve a well-chosen problem had shown real knowledge, and could keep his method private until trust was established \citep{goldstein2013routine}. Fermat's number theory survives largely in letters and in the margins of his Diophantus, published by his son in 1670; many of his results were asserted, not proved, and none was refereed \citep{goldstein2013routine}.

The most famous dispute of the period, between Newton and Leibniz over the calculus, was settled not by checking anyone's mathematics but by a committee of the Royal Society, appointed in 1712 while Newton was its president, whose report found for Newton; Leibniz was not asked for his account \citep{hall1980philosophers}. Historians now regard the two discoveries as independent. The episode is a reminder that early ``verification'' was often adjudication of priority by an institution with interests of its own.

\section{Academies and journals (1665--1900)}

The first learned journals, the \emph{Journal des sçavans} and the \emph{Philosophical Transactions}, both appeared in 1665 \citep{fyfe2022history,csiszar2018scientific}. They published at the discretion of an editor; there were no referees. In 1752 the Royal Society took control of the \emph{Transactions} and replaced its editor with a Committee of Papers that decided by vote \citep{moxham2018prehistory,rs_refereeing_blog}. Written referee reports began only in the 1830s, after William Whewell proposed that the Society commission them; the first two referees promptly disagreed \citep{baldwin2017referees,rs_refereeing_blog}.

The great nineteenth-century mathematics journals were built on editorial judgement and on the names of their authors. Crelle founded his \emph{Journal für die reine und angewandte Mathematik} in 1826, and its first volume carried several papers by Abel \citep{mactutor_crelle}; Liouville founded his \emph{Journal de mathématiques pures et appliquées} in 1836 and edited its first thirty-nine volumes, using it, among other things, to bring Galois's work to the world in 1846 \citep{mactutor_liouville}. Csiszar has shown how journals, authorship, and refereeing in this period were instruments in contests over scientific authority rather than neutral filters of quality \citep{csiszar2018scientific}.

\section{The referee becomes compulsory (1900--1990)}\label{sec:refereeing}

External refereeing became a requirement remarkably late. As Baldwin's histories show, for most of the nineteenth and early twentieth centuries referee reports were an optional tool that editors used when they saw fit; respectable venues did without them \citep{baldwin2017referees,baldwin2018scholarlykitchen}. In 1936 Einstein withdrew a paper from \emph{Physical Review} in indignation because the editor had sent it to an anonymous specialist without his permission \citep{kennefick2005einstein}. \emph{Nature} required referee reports for every paper only from 1973, and the very phrase ``peer review'', as a token of scientific legitimacy, crystallised in the politics of Cold War American science funding \citep{baldwin2015nature,baldwin2018isis}. The institution that mathematicians now treat as the guarantor of truth is, at the scale of the discipline's history, a recent arrangement.

What does a mathematics referee actually check? The honest answer, given repeatedly by people in a position to know, is: less than outsiders imagine. Nathanson wrote in the \emph{Notices of the AMS} that many, and he thought most, papers in refereed journals are not really refereed \citep{nathanson2008desperately}. Studies of the practice describe a discipline in which much knowledge is accepted on testimony, without readers or referees checking every step \citep{geist2010peer}, and in which, as Löwe has shown, measures of agreement between referees cannot reveal whether referees reliably judge correctness \citep{lowe2022measuring}. Greiffenhagen's interviews with editors of leading journals show that importance is judged first, through quick opinions, and that the checking of correctness comes later, is slow, and is never a guarantee \citep{greiffenhagen2024judging,greiffenhagen2024checking}. Grcar found that published mathematics is not notably freer of error than other fields, while its rate of published correction is very low \citep{grcar2013errors}. The American Mathematical Society's own ethical guidelines place the primary responsibility for correctness on the author, not on the journal \citep{ams_ethics}.

In 1991 the arXiv began to separate the public circulation of results, and with it priority, from journal acceptance \citep{sparc2021arxiv}. In effect it rebuilt Mersenne's network at planetary scale. It did not, by itself, decide what was correct.

\section{When refereeing was not enough}

The twentieth century supplies a sequence of cases in which the relation between refereeing, publication, and acceptance came apart.

\emph{Rigour lost and rebuilt.} The Italian school of algebraic geometry, from the 1880s to the 1930s, produced deep theorems on foundations that were later judged unsound; some claims were simply false. Severi's presumption that the rational-equivalence classes of zero-cycles on a surface form a finite-dimensional family was refuted by Mumford, who showed that the Chow group of zero-cycles is infinite-dimensional when the geometric genus is positive \citep{babbitt2012severi,mumford1968rational}. Zariski and Weil rebuilt the subject on commutative algebra \citep{weil1946foundations}. Published, refereed, celebrated work had to be re-proved from the ground up.

\emph{A proof too large to read.} The classification of the finite simple groups was announced complete in 1983. It was not: the quasithin case remained open until Aschbacher and Smith closed it in 2004 with more than twelve hundred pages \citep{solomon2001brief,aschbacher2004quasithin}. In the 1994 responses to Jaffe and Quinn, Mac Lane observed in passing that part of the classification was not yet published \citep{atiyah1994responses}. The original proof is spread over hundreds of articles by about a hundred authors; a second-generation proof is still being written \citep{solomon2001brief}. No referee ever read the whole.

\emph{A referee finds the gap.} When Wiles announced a proof of Fermat's Last Theorem in June 1993, it was the refereeing process, through Nick Katz's questions, that exposed a gap; a year later Taylor and Wiles repaired it, and the two papers filled an issue of the \emph{Annals} in May 1995 \citep{faltings1995proof,wiles1995modular,taylorwiles1995ring}. This is refereeing at its best, and it is worth noticing how exceptional the circumstances were: the most important result in decades, read by the best possible experts, with the whole world watching.

\emph{Accepted without a journal.} Perelman posted three preprints on the arXiv in 2002 and 2003 and never submitted them anywhere \citep{perelman2002entropy,perelman2003ricci,perelman2003finite}. Acceptance came through the work of others: teams who wrote long expository accounts that checked and filled in his arguments, one of which became the subject of a dispute about credit \citep{kleiner2008notes,morgan2007ricci,cao2006complete,nasar2006manifold}. He declined the Fields Medal in 2006 and the Clay Millennium Prize in 2010 \citep{nasar2006manifold,cmi2010annual}. Correctness was established by a community, not by a journal; the journal system's other function, the allocation of honours, he refused.

\emph{Published but not accepted.} In 2012 Mochizuki posted four papers on inter-universal Teichmüller theory claiming a proof of the $abc$ conjecture. In 2018 Scholze and Stix reported a gap at Corollary~3.12 which, in their view, no small modification could repair \citep{scholzestix2018why}; Mochizuki rejected their reading \citep{mochizuki2018report}. In 2021 the papers were published in \emph{Publications of the Research Institute for Mathematical Sciences}, the journal of Mochizuki's own institute, of which he is chief editor; the decision was taken by a special editorial committee, led by Kashiwara and Tamagawa as co-chief editors, from which Mochizuki was excluded \citep{mochizuki2021iut,castelvecchi2020mathematical,kyotonp2020mochizuki}. The proof has been refereed and published. It has not been accepted by the community.

\emph{The doubter turned builder.} Voevodsky found, years after publication, that a key lemma in one of his own important papers was wrong, and that a well-known paper he had written in 1989 (published in 1991) claimed something false; the errors had survived because no one checks in detail the hard technical arguments of a trusted author. He concluded that only computers could guarantee correctness at scale and spent his last years on foundations for machine-checked mathematics \citep{voevodsky2014origins,voevodsky2014univalent}.

\section{The arguments about what proof is for}

The practical crises produced a theoretical debate. De Millo, Lipton, and Perlis argued in 1979 that confidence in a theorem is the product of a social process (reading, doubting, simplifying, teaching, using) rather than of formal derivation, and concluded that formal verification of programs could never do what proofs do for mathematics \citep{demillo1979social}. In 1993 Jaffe and Quinn proposed that speculative results be clearly labelled and distinguished from rigorous ones \citep{jaffequinn1993theoretical}. Thurston replied that what mathematicians produce is human understanding, and that the reliability of mathematics does not come primarily from formal checking \citep{thurston1994proof}; Atiyah objected to the authoritarian tone of the proposal \citep{atiyah1994responses}. These essays define the positions that the events of 2026 put to the test. De Millo, Lipton, and Perlis were right that acceptance is social. They were wrong about where the social process would end up.

\section{The machine enters (1976--2026)}

In 1976 Appel and Haken proved the four colour theorem by reducing it to a finite set of configurations checked by computer, a check no human could perform by hand \citep{appel1977everyI,appel1977everyII}. Tymoczko argued that accepting it admitted an empirical element into mathematics \citep{tymoczko1979four}. Robertson, Sanders, Seymour, and Thomas produced a simpler computer proof in 1997 \citep{robertson1997four}, and in 2005 Gonthier completed a proof checked down to the small kernel of the Coq proof assistant \citep{gonthier2005computer,gonthier2008formal}. The objection ``can we trust the program?'' was answered by moving from special-purpose code to a general logical kernel.

In 1998 Hales announced a proof of the Kepler conjecture, some 250 pages and three gigabytes of computation. After years of work, a panel of referees for the \emph{Annals} declared itself, as reported, only ``99\% certain'' and unable to check all the computations \citep{szpiro2003kepler}. While the review was stalled, in 2003, Hales launched Flyspeck, a project to formalise the entire proof; the main paper appeared in 2005 with the referees' caveat \citep{hales2005proof}, and Flyspeck was completed in August 2014 and published in 2017 \citep{hales2017formal,phys2014flyspeck}. Here, for the first time, a proof that refereeing could not certify was certified by machine. The Feit--Thompson odd order theorem, a cornerstone of the classification, was fully formalised in Coq in 2012 \citep{gonthier2013oddorder}.

The Lean proof assistant, begun by de Moura in 2013, and its community library Mathlib, became the working tool of a generation of mathematicians \citep{demoura2021lean4}. In December 2020 Scholze challenged the community to verify a central theorem of condensed mathematics that he himself was unsure of; the Liquid Tensor Experiment completed it in July 2022 \citep{lte2022completion}. In November 2023 Tao and collaborators began formalising the polynomial Freiman--Ruzsa conjecture days after its proof appeared, and finished on 5 December, organised by Massot's blueprints \citep{tao2023pfr,massot_leanblueprint}. In 2024 Buzzard began a funded five-year project to formalise Fermat's Last Theorem \citep{buzzard2024flt}; in September 2026 Anthropic announced that Claude had completed a formal proof of the theorem, largely autonomously, in eleven days and more than thirteen million lines \citep{anthropic2026flt}. Systems such as AlphaProof had already shown that AI could produce machine-checked proofs of competition problems \citep{deepmind2025alphaproof}.

Over these fifty years the question asked of machine proofs changed. In 1976 it was whether a computation could be trusted. By 2026 the kernel's verdict was trusted, and the question had become whether the formal statement says what it is supposed to say. Buzzard's four criteria for trusting a Lean proof, published eight days before the Riemann Hypothesis was announced, begin with exactly that question \citep{buzzard2026criteria}.

\section{What the history shows}

A caution is in order. Several of the early episodes, the cubic and the calculus above all, were disputes over priority rather than over correctness, and the history as a whole can be read as a long drift rather than a series of breaks. I accept that reading. The claim of this paper is not that 2026 overturned a stable regime, but that it completed a movement that had been under way for half a century, and did so in a single, visible event. With that caution, three things follow from this review. First, each regime was a contingent technology of trust: public contest, correspondence, academy, editor, referee, community. None was a timeless standard. Second, refereeing, the regime of the twentieth century, never reliably delivered what the public believed it delivered; its practitioners have said so. Third, the most difficult cases of the last fifty years, from Kepler to the Liquid Tensor Experiment, were resolved not by more refereeing but by formalisation. The machine did not arrive as a rival to a healthy institution. It arrived to do the job the institution could not finish for the largest and hardest proofs.

It is also worth noticing a return. The challenge problems of Mersenne's circle were an objective, author-blind test: either you could solve the problem or you could not. A formal statement in Mathlib with a kernel waiting to check is a challenge in the same spirit, posed to anyone in the world.

%% file: sections/20_part2.tex
\part*{Part II. November 2026}
\addcontentsline{toc}{part}{Part II. November 2026}

\section{A proof released as code}\label{sec:rh}

On 20 November 2026 Levent Alpöge posted a four-page note and a Lean archive of 175,187 files and 17,374,215 lines \citep{alpoge2026remark,sutherland2026verification}. The note says that he believes he has, working with Claude, found a proof of the Riemann Hypothesis and of the Generalized Riemann Hypothesis for Dirichlet $L$-functions, and that the arguments have been formalised without assumptions. It does not contain the proof. Alpöge explains why: he had not yet mastered the large body of mathematics surrounding the argument, and he judged it inappropriate to post an enormous unreadable document; so he released only Lean \citep{alpoge2026remark}.

What followed took about a day. The community opened a repository to clean up the archive. Independent participants rebuilt it and replayed every module through Lean's kernel; Andrew Sutherland compiled all 175,187 modules, listed the axioms of the five final theorems, and ran an independently written kernel over the exported proof \citep{sutherland2026verification}; Michael Rubinstein ran the same checks and reported no errors \citep{rubinstein2026guide}. Sutherland is also the editor-in-chief of a number theory journal: a gatekeeper of the old system was re-running the new one on his own machine. On 21 November the Lean FRO and the Mathlib Initiative issued a joint statement: they certified that both the statement and the proof are correct, and that all four of Kevin Buzzard's criteria for trusting a Lean proof were met \citep{leanfro2026statement,buzzard2026criteria}. The statement proved is Mathlib's own \texttt{RiemannHypothesis}; it was compared against the proved theorem with two independent tools; and the exported proof was accepted by Lean's kernel and by several independently written checkers, two of which come with a formal proof that they cannot accept a false proof \citep{leanfro2026statement}.

The proof then began to shrink. On 25 November a rewrite proving the same five theorems appeared, with a compact edition of 112,827 lines, roughly one hundred and fifty times shorter than the original (about 76,000 lines of code) \citep{sernau2026grhlean,naskrecki2026riemannrev}; independent checks of it, under six kernels, finished the next day \citep{rubinstein2026guide}. Human-readable accounts began to appear: a 265-page working guide specialised to the zeta function, which reduces the central inequality to thirteen hypotheses it does not itself prove \citep{rubinstein2026guide}, and a unit-by-unit reading of the compact edition made by AI agents and not yet refereed by a human \citep{naskrecki2026riemannrev}.

Two features of this sequence matter for the argument. First, certification did not wait for a complete human-readable reconstruction of the proof. Alpöge's note explains what he set out to transfer from Stepanov's method over function fields and why the chain length $\sqrt X$ struck him \citep{alpoge2026remark}; but no one had yet written the argument out for human readers. Second, certification was not granted by an editor or a referee. It was granted by a public process whose every step could be repeated by anyone with a computer: compile, list the axioms, compare the statement, replay the kernel. The ``referees'' were the people who wrote the statement years earlier and the people who ran the checks on 20--21 November, and their ``reports'' are public logs that anyone can re-run. Acceptance in the fuller sense, by mathematicians who have read and absorbed the argument, has not yet happened; that is the subject of Part~III.

\section{A mirror pair}\label{sec:mirror}

Set beside this the most discussed contested proof of the previous decade. Mochizuki's claimed proof of the $abc$ conjecture, posted in 2012, was published in 2021 in the journal of his own institute, of which he is chief editor (a special committee from which he was excluded made the decision), after Scholze and Stix had reported a gap that in their view no small modification could repair (see Part~I) \citep{mochizuki2021iut,scholzestix2018why}. It has been refereed and published. It has not been accepted.

The two cases are mirror images. One proof passed peer review and was not accepted. The other never went near peer review and was certified correct in a day. Between them sits Perelman's proof of the Poincaré conjecture, posted only on the arXiv and accepted after years of expository work by others, without its author ever submitting it to a journal \citep{kleiner2008notes,nasar2006manifold}. The sequence journal $\to$ community $\to$ machine is the movement of authority that Part~I traces over three and a half centuries, compressed into three examples.

The pair is not a controlled experiment. Mochizuki's argument has never been formalised, so the two proofs did not face the same test and diverge; and Perelman's case shows that a human community can establish correctness without a machine, given years and the right people. What the pair shows is narrower and, I think, more important: in 2026 the authority of a published, refereed paper and the authority of a checked formal proof came visibly apart, and the community followed the second.

I do not mean that Mochizuki's work is wrong or that Alpöge's is understood. I mean that, in these cases, the question ``has it been refereed?'' told us less than the question ``does it compile against the agreed statement?''.

\section{My own case: where the proof stopped}\label{sec:mycase}

The prologue described the letter. Here is what happened next, stated as plainly as I can, because a paper that criticises refereeing must not hide its author's failures.

The submitted paper argued that no $3\times 3$ magic square of distinct squares exists. Following the reviewer's advice, I rebuilt the argument in Lean, under a strict rule: a result counts only if its proof depends on Lean's three standard axioms and nothing else. Large parts closed. Whole families of cases were eliminated, some through the arithmetic of a particular elliptic curve of rank zero, and the formal statements were checked against computer-algebra calculations before their proofs were written. But one step would not close. Lean checked the reductions. Formalising the final step forced me to state, as a separate claim, what my manuscript had passed over. Once I could see that claim on its own, I could also see that the manuscript had supplied no proof of it. Concretely, the argument is an induction on $k$, the number of distinct primes $\equiv 1 \pmod 4$ dividing the entries. Lean closed the cases $k\le 2$, the case $k=2$ through the arithmetic of the elliptic curve with LMFDB label 48.a3, which has rank zero. The general inductive step to larger $k$ did not close: my manuscript had carried it by an argument about discriminants that are not perfect squares as polynomials. Not being a square as a polynomial is weaker than never taking square values: $t^2+1$ is not a square, yet it equals $1$ at $t=0$. Formalising the step forced me to state, as a separate hypothesis, the claim it actually needed. That hypothesis is the remaining condition on which my nonexistence argument depended, and the sixty pages contained no proof of it. Lean did not find the gap the way a referee finds an error. It verified, with the standard axioms alone, that the remaining claim implies the theorem, because the claim enters as a hypothesis and not as an axiom. What it forced was that the hypothesis be written down. Locating where the sixty pages had stepped over it was then my job.

The reviewer's judgement, that the paper was very unlikely to be correct, was therefore right. I want to be exact about what kind of rightness it was. It was a forecast, based on the history of failed attempts. It was not a finding. The reviewer did not know where the error was, and neither did I. But the reviewer was right in a second and deeper sense. Not knowing the answer, the reviewer named a method by which I could find it out, and the method was sound whichever way it came out: had the formalisation closed, the advice would have produced a checked proof; because it did not, it showed me exactly what I had not proved. Formalisation located the gap, and would have located it no matter who had written the paper.

That is the whole argument of this essay in one case. The reviewer and the machine were both right, in different senses, and the difference is the point. The reviewer issued a forecast: a judgement about plausibility, drawn from the history of submissions of this kind. Lean returned a conditional affirmation: the theorem holds if the remaining condition does, and nothing in the check depended on who had written the proof. In November the same kind of check accepted a proof of the Riemann Hypothesis from a mathematician working with an AI, before any human had written it out. The reviewer's prescription, to formalise the result in Lean, could easily be read as a polite way of closing a door. The reviewer seems to have meant it literally, adding that AI coding agents would make it feasible. It turned out to describe the new door.

Set beside the Riemann Hypothesis, the case shows one more thing: formalisation changes where the next person can begin. Before it, anyone wishing to judge my paper had to start at page one of sixty; after it, they can start from a single stated hypothesis and ask why it should hold. After 21 November, mathematicians wishing to understand Alpöge's proof could start from a checked theorem and ask why a particular device works and whether it carries over to other problems \citep{sahay2026polymath}. In neither case is the change best measured by referees saved. What changed is where human attention goes.

%% file: sections/30_part3.tex
\part*{Part III. What changes, and what remains}
\addcontentsline{toc}{part}{Part III. What changes, and what remains}

\section{Five things a referee was asked to do}\label{sec:unbundling}

Refereeing has always been a bundle. When an editor sends a mathematics paper to a referee, the referee is implicitly asked five different questions, which the institution has never needed to separate:
\begin{enumerate}
\item \textbf{Correctness.} Are the arguments free of gaps?
\item \textbf{Fidelity.} Does the paper prove what it says it proves? Do the theorems mean what the introduction claims they mean?
\item \textbf{Significance.} Does the result matter? Is it new? Is it worth the journal's pages and the readers' attention?
\item \textbf{Understanding.} Can other mathematicians read it, check it, and use it? Is the exposition adequate?
\item \textbf{Credit and allocation.} By extension, through publication: who gets jobs, grants, prizes, and attention.
\end{enumerate}
These functions travel together because they have been performed by a few readers of one manuscript, even where leading journals had begun to separate a quick judgement of importance from a slower check of correctness \citep{greiffenhagen2024judging}. The bundle has costs that the history in Part~I makes plain. Correctness, the function the public believes refereeing performs, is the one referees are least able to guarantee for long proofs; the reviewer who declined my paper said so directly, and so have editors and historians of the discipline (\S\ref{sec:refereeing}). Significance and credit, the functions that matter most to careers, are the ones most exposed to prestige, network, and prior belief.

A proof assistant performs exactly one of the five, and part of a second.

It performs (1) completely. A kernel either accepts a proof term or it does not, and the axioms on which an accepted proof depends can be listed mechanically. ``The proof uses only the three standard axioms'' is a fact anyone can check in minutes; it is not an opinion about the author.

It performs (2) only partly. The kernel certifies that \emph{some} statement has been proved. Whether that statement is the one the authors claim, whether the formal definition of ``the Riemann Hypothesis'' really is the Riemann Hypothesis, is a question about the relation between symbols and intentions, and no machine can settle it on its own. Buzzard's first criterion for trusting a Lean proof is precisely this one \citep{buzzard2026criteria}. In November 2026 it was settled in a way that is worth noticing: the statement Alpöge proved was Mathlib's own \texttt{RiemannHypothesis}, a definition written and vetted by experts long before the proof existed, and the community confirmed with two independent tools that the theorem proved is exactly that definition \citep{leanfro2026statement}. Fidelity was settled in two steps that are worth keeping apart. The meaning of the problem was fixed \emph{before} the proof, by people, in public. That the new proof reached exactly that meaning was confirmed \emph{after} it, again by people, with tools. Sutherland went further, connecting the result through separately formalised equivalences of Robin and Lagarias to arithmetic inequalities that do not mention zeros at all \citep{sutherland2026verification}. Fixing what a problem means and confirming that a newly arrived proof reaches it are different tasks, in time and in mathematics.

It performs (3) not at all, but for a famous problem it does not need to. Nobody needs a referee to tell them that the Riemann Hypothesis matters.

It performs (4) not at all, and here the November proof is a striking case: certification did not wait for a human-readable reconstruction. Alpöge wrote frankly that he had not yet mastered the argument he released \citep{alpoge2026remark}. The first human-readable guides that followed reduced the central inequality to a list of hypotheses they did not themselves prove \citep{rubinstein2026guide}. For the first time, a statement many regard as the central problem of number theory was certified, against an agreed formal statement, before a complete human-readable proof existed. Three days before the release, Vallverdú had argued that the warrant for a mathematical result belongs to the components of a ``warrant architecture'' rather than to a proof as a whole \citep{vallverdu2026warrant}; November showed such an architecture at work, with the components held by different people and machines.

And it performs (5) not at all. Who deserves credit for a proof in which, by the author's own account, a machine proposed the decisive choice? \citep{alpoge2026remark}.

\section{Review moves upstream: the statement is the review}\label{sec:upstream}

If the kernel decides (1), and (2) is decided by a statement written in advance, then the act that matters most in judging a claimed proof has moved. It is no longer the reading of the proof after it arrives. It is the writing of the statement before anyone tries.

This changes who the referees are. In the traditional sense, Alpöge's proof had no journal referee; in a new sense, it had referees all along. The people who wrote and checked Mathlib's definition of the zeta function and of the Riemann Hypothesis, years before Alpöge's work, were the effective referees of his theorem. They could not have known it. Their review was real mathematical scrutiny. Buzzard and Ellenberg had noticed that an early formulation depended on the value that Lean assigns to the zeta function at its pole $s=1$, a value nobody had chosen: had it been zero, the stated hypothesis would have been trivially false. Loeffler and Stoll showed that the value is nonzero, about $-0.98$, and the current statement excludes $s=1$ explicitly \citep{loeffler2025zeta}. After Alpöge's release, the community checked that his proof established this vetted statement \citep{leanfro2026statement}.

It also suggests what the institution of the future could look like for a particular class of problems: famous, canonical, and statable in an existing library, as the Riemann Hypothesis was. Most mathematics is not like this; for ordinary papers the statement is not registered in advance, significance needs expert judgement, and formalisation remains expensive. For the canonical problems, though, the outline is clear: a public registry of formally stated open problems, each with a challenge file that imports only the standard library; anyone, human or machine, may submit a proof; a kernel decides; a comparison tool confirms the submitted theorem matches the registered statement. Such a registry already exists in embryo: Formal Conjectures collects open problems stated in Lean \citep{firsching2026formal}, and among them is the $3\times 3$ magic square of squares \citep{formalconjectures_magic}. The problem I failed to solve and the Riemann Hypothesis, which Mathlib had stated, stand on the same kind of public foundation. The Clay Institute's rules, which require publication in a qualifying outlet, two years of waiting, and ``general acceptance'' \citep{cmi2018rules}, were written for a world in which acceptance had to be slowly accumulated because correctness could not be checked directly. That world has ended for any problem that can be registered in this way. A prize committee would still need to decide what the statement should be, and that it can decide in advance.

There is a risk in this, and it is the same risk that refereeing always carried, moved earlier: a badly chosen statement will be proved correctly and mean nothing. The statement-writers inherit the referee's responsibility in a purer form. They can no longer hide behind the difficulty of reading the proof.

\section{What remains human}\label{sec:human}

Three things remain, and none of them is the kernel's kind of checking.

\emph{Asking.} Deciding which statements deserve to be formalised and attacked is the real judgement of significance. In a world where correct proofs can be produced in quantity, the scarce resource is not correctness but well-chosen questions.

\emph{Understanding.} Thurston argued in 1994 that mathematicians do not want proofs so much as understanding, and that a proof is valuable mainly for what it makes people understand \citep{thurston1994proof}. The November proof puts his claim to its hardest test. Its truth is settled; a full mathematical exposition is still being worked out, in preliminary notes such as Zhang's \citep{zhang2026preliminary}. I will call the gap between the two an \emph{understanding debt}. The debt is not owed in order to establish truth. It is owed because methods only travel through minds: unless someone understands why a chain of theta lifts of length $\sqrt X$ works, the idea cannot be carried to the next problem. Paying this debt is now the central scholarly task, and it is work that the present credit system barely rewards. A ``guide'' to someone else's proof is not a theorem.\footnote{My own small contribution is a public guide for non-specialists, Riemann Atlas (\url{https://riemannatlas.org}).} Yet the work has begun without waiting for incentives: three days after the certification, Sahay, Sawin and Silberman proposed a Polymath project, one of whose stated goals is a human-readable proof of the Riemann Hypothesis and its generalisation \citep{sahay2026polymath}.

\emph{Credit.} The machine that proposed the decisive step has no career. The person who released the proof says he does not fully understand it. The people who will explain it will not have discovered it. Our vocabulary of authorship assumes these roles coincide. They no longer do.

\section{The outsider's view}\label{sec:outsider}

I write as an outsider, and I want to say two things that are easier to say from outside.

The first is that refereeing, as practised on unsolicited claims to famous results, was never a test of correctness. It was a filter based on prior probability. The reviewer of my paper said as much, honestly, and was right on the facts of my case. But the judgement was about the kind of submission, drawn from the record of earlier ones, and was made without checking this one. Desk rejections of my other papers in 2026 sometimes cited readability, scope, or a suspicion that the text was written with AI; some of those criticisms were fair. None of them was a check. A proof checker is indifferent to who the author is. It showed where my magic-square proof stopped as indifferently as it accepted Alpöge's proof of the Riemann Hypothesis.

The second point concerns my own case, and I want to keep it there. For years, people outside the academy who believed they had solved famous problems could say that the gate was unfair, that no one would read them. In my case that explanation is gone. My magic-square proof did not fail because I am a psychiatrist; it failed because there was a step I had not proved, and the kernel would have said so to anyone. I cannot honestly claim more. The kernel's verdict on a given proof does not depend on its author, but whose files are checked at all, read, and built upon is still decided by people. Had the same Riemann Hypothesis files come from an unknown author, the kernel would have answered the same; whether dozens of people would have given their machines to it by the next morning, this case does not tell us.

\section{Limits}\label{sec:limits}

The argument has a boundary, and it should be drawn clearly.

It applies to claims that can be stated formally. Most of science is not like that. In medicine, psychology, or the social sciences, correctness is not a property of an argument alone; it depends on data, measurement, and the world. There the analogue of a proof checker is reproducible analysis on open data: a machine can re-run what was done, and some work toward this is under way, including my own small attempt to re-run official Japanese statistics \citep{urazuke2026}. But a machine cannot decide whether a clinical judgement was wise or whether a measurement captured what it claimed to. In those fields refereeing will keep more of its bundle.

Even within mathematics, fidelity remains a human judgement, and it is the place where formal verification can most easily be misused. A proof of the wrong statement is perfectly checkable. The November proof was trustworthy in part because its statement had been fixed by others in advance; a proof of a statement chosen by its own author deserves more scrutiny, not less.

And understanding cannot be automated in the sense that matters here. An AI may well write an excellent exposition of the November proof. But the debt is paid only when people can use the ideas, and that is a fact about people.

%% file: sections/40_coda.tex
\section*{Coda: an unrefereed paper}

This paper has not been refereed and will not be submitted to a journal. That is not a protest; it seems the most consistent way to publish an argument of this kind. Its claims about history are checkable against the sources cited; its claims about November 2026 are checkable against public logs; its argument is offered to readers, not to an editor.

There is a self-reference here that I enjoy. Gödel showed that a sufficiently strong formal system cannot certify its own consistency from inside. Refereeing, too, could never certify its own reliability: the only check on referees was more referees. The machine has not escaped Gödel; Lean's kernel rests on axioms and on software that people must trust. But it has moved the point of trust from a person's judgement of another person to a small, public, repeatable procedure. That is a smaller and better place to put one's faith.

What remains is to understand what we now know to be true. That work has no machine, and no deadline. It is the part of mathematics that was always the point.

\section*{Note on the use of AI}

The literature search, drafting, and checking of this paper were done with the help of Claude (Anthropic) running as research and coding agents, and the drafts were reviewed by other language models. The argument, the judgements, and any errors are mine.